\documentclass[pre,aps]{revtex4}

\usepackage{graphicx}
\usepackage{amsmath}

\newcommand{\beq}{\begin{equation}}
\newcommand{\eeq}{\end{equation}}

\begin{document}

\title{
Fluctuations of entropy production in the isokinetic ensemble
}
\author{
F.~Zamponi$^{1}$\footnote{e-mail: francesco.zamponi@phys.uniroma1.it},
G.~Ruocco$^{1}$, and L.~Angelani$^{1,2}$
        }
\affiliation{
         $^1${Dipartimento di Fisica and INFM, Universit\`a di Roma
         {\em La Sapienza}, P. A. Moro 2, 00185 Roma, Italy}
        }
\affiliation{
        $^2${INFM - Center for Statistical Mechanics and Complexity,
        Universit\`a di Roma {\em La Sapienza}, P. A. Moro 2, 00185 Roma,
Italy}
        }

\date{\today}
\begin{abstract}
We discuss the microscopic definition of entropy production rate in a
model of a dissipative system: a sheared fluid in which the kinetic
energy is kept constant via a Gaussian thermostat.  The total phase
space contraction rate is the sum of two statistically independent
contributions: the first one is due to the work of the conservative
forces, is independent of the driving force and does not vanish at
zero drive, making the system non-conservative also in equilibrium.
The second is due to the work of the dissipative forces, and is
responsible for the average entropy production; the distribution of
its fluctuations is found to verify the Fluctuation Relation of
Gallavotti and Cohen.  The distribution of the fluctuations of the
total phase space contraction rate also verify the Fluctuation
Relation.  It is compared with the same quantity calculated in the
isoenergetic ensemble: we find that the two ensembles are equivalent,
as conjectured by Gallavotti. Finally, we discuss the implication of
our results for experiments trying to verify the validity of the FR.
\end{abstract}
\maketitle

%%%%%%%%%%%%%%%  TEXT  %%%%%%%%%%%%%%%%


\section{Introduction}

A very important concept in the theory of out-of-equilibrium
stationary states induced by the application of a driving force
(temperature or velocity gradients, electric fields, etc.)  to a
system in contact with a thermal bath is that of {\it entropy
production rate} \cite{gal1,gal2}.  In usual nonequilibrium
thermodynamics it is defined as the power dissipated by the driving
force divided by the temperature of the bath \cite{evans}: \beq
\label{DEFsigma}
\dot{\sigma}(t)=\frac{1}{T} \sum_i J_i(t) X_i(t) \ .  \eeq Here
$X_i(t)$ is the driving force (e.g. the electrical field or the
temperature gradient) and $J_i(t)$ its conjugated flux (the electrical
current or the heat flux respectively).

The Probability Distribution Function (PDF) of the total entropy production
over a time $\tau$ is defined as
\beq
P_\tau(\sigma)=P\left( \int_0^\tau dt \ \dot{\sigma}(t)=\sigma \right) \ .
\eeq
If we set $k_B=1$, $\dot\sigma(t)$ has the dimension of an inverse time and
$\sigma$ is dimensionless.
Then, in the limit of large $\tau$, the function $P_\tau(\sigma)$ is expected
to verify the {\it Fluctuation Relation} (FR) in the following form:
\beq
\label{FT}
P_\tau(-\sigma)=e^{-\sigma} \ P_\tau(\sigma) \ , \eeq if
$|\sigma/\langle \sigma \rangle| < C$, being $C$ a positive constant.
Having fixed $C$, the correction at finite $\tau$ is of order one in
the exponent of the right side of Eq.~\ref{FT} (while $\sigma$ is of
order $\tau$ if $\gamma \neq 0$).  The FR states that the probability
to observe a negative entropy production (over a large enough time
interval) is exponentially smaller than the probability to observe the
same value with positive sign.  This relation has been first observed
numerically in a sheared fluid \cite{ECM} and subsequently proven to
hold for {\it reversible} systems by Gallavotti and Cohen under
the {\it chaotic hypothesis}, a strong chaoticity assumption for the
dynamics of the system, leading to the demonstration of the Fluctuation
Theorem (FT) \cite{GC}.
Gallavotti then showed that in the limit of small driving forces the
FT implies the usual Green-Kubo relations and the
Fluctuation-Dissipation Theorem \cite{GAL}, thus clarifying the deep
physical significance of Eq.~\ref{FT}.  Despite the hypothesis of the
FT are strictly verified only for very special systems, the FR has
been shown to be valid for a large class of dissipative systems in
very different conditions \cite{BGG,BCL,Kurchan,LebSpo,vulpiani,GP}.
Some experimental attempts have also been done in order to check its
validity for real systems \cite{ciliberto,goldburg,menon}.

The verification of Eq.~\ref{FT} in microscopic models for dissipative
systems requires a microscopic definition of $\dot{\sigma}(t)$.  In
the case of conservative models, in which the phase space volume is
conserved in absence of drive, the entropy production rate has been
identified at the microscopic level with the phase space contraction
rate \cite{ECM,BGG,BCL,GP}.  One can ask if the same identification
holds for models in which the phase space volume is not conserved even
in absence of the driving force.

It has been conjectured by Gallavotti that, if the entropy production rate
is properly defined, equivalence
of ensembles should hold, in the sense that the FR must hold for a subsystem of
the (big) dynamical system
under consideration, at least for a large class of thermostatting mechanisms,
including irreversible and stochastic ones \cite{gal1,gal2,gal_stat,gal_fluid}.

In this paper we will discuss two different models of thermostat, both
{\it mechanical} and {\it reversible} (then, we will not discuss the problem
of equivalence between reversible and irreversible thermostats), defining two
different ensembles:
the first one is constructed in order so that the total phase space contraction
rate vanishes in equilibrium,
while in the second one fluctuations of the total phase space contraction rate
are present even
in the limit of
zero driving force, thus making the system non-conservative also in
equilibrium.
We will then discuss the equivalence of these ensemble and the possibility of
identifying the entropy production
rate with the phase space contraction rate when the latter is not vanishing in
equilibrium.

\section{Two models of a mechanical thermostat}

We consider a well known microscopic model for a
sheared fluid defined by the SLLOD equations \cite{evans}:
\beq
\begin{cases}
&\dot{q}_i = \frac{p_i}{m} + \gamma y_i \hat{x} \ , \\
&\dot{p}_i = F_i - \gamma p_{yi} \hat{x} - \alpha(p,q) p_i \ ,
\end{cases}
\eeq
where $F_i$ are conservative forces, $F_i=-\partial_{q_i} V(q)$. The terms
proportional
to $\gamma$ (the shear rate) impose to the liquid a flow
along the $x$ direction with a gradient velocity field along the $y$ axis.
In this model the driving force is the velocity gradient $\gamma$,
and its conjugated flux is the $yx$ component of the stress tensor, $P_{yx}$.
The function $\alpha(p,q)$ is a Gaussian thermostat, and can be defined in
order to conserve either the total energy $H(p,q)=\sum_i \frac{p_i^2}{2m} +
V(q)$ or the kinetic energy alone.
The total phase space contraction rate for this system is given by:
\beq
\dot{\sigma}(p,q)=-\sum_i \left( \frac{\partial \dot{q}_i}{\partial q_i}
+\frac{\partial \dot{p}_i}{\partial p_i} \right)
= 3N\alpha(p,q) + \sum_i \frac{\partial \alpha}{\partial p_i} p_i \ .
\eeq
The second term is of order one in the case we will discuss,
and will be neglected with respect to the first term.

\subsubsection{Isoenergetic (or microcanonical) ensemble}

Imposing the constraint $dH/dt=0$, one gets the following expression for
$\alpha$:
\beq
\alpha_H(p,q)=-\gamma \frac{\sum_i ( p_{xi} p_{yi} + m F_{xi} y_i )}{\sum_i
p^2_i} \ .
\eeq
The expression for the phase space contraction rate is then:
\beq
\label{sigmagamma}
\dot{\sigma}_{H}(p,q) = 3N\alpha_H(p,q) = -\frac{\gamma P_{yx}(p,q)}{T(p)}
\equiv \dot{\sigma}_\gamma(p,q) \ ,
\eeq
where $T(p)=\frac{1}{3N}\sum_i \frac{p^2_i}{m}$ and $P_{yx}(p,q)=\sum_i (
p_{xi} p_{yi}/m + F_{xi} y_i)$ is the
microscopic expression of the $yx$ component of the stress tensor \cite{evans}.
Note that in this case the phase space contraction rate is exactly equal to the
microscopic
expression of Eq.~\ref{DEFsigma}.
From Eq.~\ref{sigmagamma} we see that in the isoenergetic ensemble
{\it the phase space volume is conserved in equilibrium} as $\dot{\sigma}_H$ is
vanishing for $\gamma=0$.
The behavior of the fluctuations of $\dot{\sigma}_H$ has been discussed in
\cite{ECM} where the validity of
the FR for $P_\tau(\sigma_H)$ was observed for the first time.

\subsubsection{Isokinetic ensemble}

If the temperature $T(p)$ has to be conserved instead of the total energy,
one obtains the following expression for $\alpha$:
\beq
\alpha_T(p,q)=\frac{\sum_i p_i F_i - \gamma \sum_i p_{xi} p_{yi}}{\sum_i p^2_i}
=
\alpha_H(p,q) + \frac{\sum_i m \dot{q}_i F_i}{\sum_i p^2_i} \ ,
\eeq
and the total phase space contraction rate is given by
\beq
\label{sigmaisoK}
\dot{\sigma}_T(p,q) = -\frac{\gamma P_{yx}(p,q)}{T(p)} + \frac{\sum_i \dot{q}_i
F_i}{T(p)} =
\dot{\sigma}_\gamma(p,q) + \dot{\sigma}_c(p,q) \ .
\eeq
From the previous expression one sees that the total phase space volume
contraction rate in the
isokinetic ensemble is the sum of two different contributions:
the first one ($\dot{\sigma}_\gamma$) is due to the work of the
{\it dissipative} forces and has the same microscopic expression as in the
isoenergetic
ensemble (see Eq.~\ref{sigmagamma}); the second one ($\dot{\sigma}_c$) is the
power dissipated
by the conservative forces divided by the temperature.
It is easy to see that it the second term can be also written as
$\dot{\sigma}_c(p,q)=-T^{-1}(p)\frac{dV(q)}{dt}$;
thus, it has zero average (because the total potential energy is constant in
average).
This term is present also in equilibrium ($\gamma=0$): then, in the isokinetic
ensemble,
{\it the phase space volume is not conserved also in equilibrium}, at variance
to what
happens in the isoenergetic ensemble. However, the two ensembles are known to
be equivalent in
equilibrium \cite{evans}.

\subsubsection{Ensemble equivalence in nonequilibrium}

Having defined these two model of thermostat, two questions naturally arise: \\
{\it 1)} One can ask if the proper definition of $\dot{\sigma}(t)$ (i.e.
the one that verifies the FR) in the isokinetic ensemble
is given by the total phase space contraction rate $\dot{\sigma}_T$ or by
$\dot{\sigma}_\gamma$ alone; \\
{\it 2)} Once a definition for the entropy production rate in the isokinetic
ensemble has been chosen,
one can compare the distribution of its fluctuation with the same quantity
calculated in the
isoenergetic ensemble, thus verifying if some kind of nonequilibrium ensemble
equivalence holds \cite{gal1,gal2}.

To address this points, we will check numerically the validity of the following
statements: \\
{\it i)} $\dot{\sigma}_\gamma$ and $\dot{\sigma}_c$ defined in
Eq.~\ref{sigmaisoK} are statistically independent
in the isokinetic ensemble; \\
{\it ii)} the PDF of $\dot{\sigma}_c$ is $\gamma$-independent in the range
of $\gamma$ explored, i.e. the fluctuations of the power
dissipated by the conservative forces are closely the same in and 
out of equilibrium; \\
{\it iii)} the PDF of $\dot{\sigma}_\gamma$ is the same in the isokinetic and
in the isoenergetic
ensemble for any value of $\tau$, therefore $P_\tau(\sigma_\gamma)$ verifies
the FR in both ensembles; \\
{\it iv)} the large $\tau$ limit for $P_\tau(\sigma_\gamma)$ (condition for the
validity of the FR) is attained for
$\tau \gtrsim 10^2 \tau_\alpha$, being $\tau_\alpha$ the decay time of the
stress correlation function; \\
{\it v)} in the isokinetic ensemble, at large $\tau$, the fluctuations of
$\dot{\sigma}_T$ are completely dominated
by the dissipative part $\dot{\sigma}_\gamma$, and $P_\tau(\sigma_T)$ tends to
$P_\tau(\sigma_\gamma)$: thus,
$P_\tau(\sigma_T)$ also verifies the FR. However, this happens for $\tau$
values ($ \gtrsim 10^4 \tau_\alpha$)
much greater than the ones needed to observe the FR for
$P_\tau(\sigma_\gamma)$. \\
From the above statements it follows that $\dot{\sigma}_c$ acts as a
``noise'' superimposed to
$\dot{\sigma}_\gamma$. It does not contribute to the average entropy
production, and contributes to its fluctuations
only for small $\tau$.

\section{Details of the simulation}

The investigated system is a binary mixture of $N$=66 particles (33 type A and
33 type B) of equal mass
$m$ interacting via a soft sphere pair potential
$V_{\alpha \beta}(r)=\epsilon \left( \frac{\sigma_\alpha+\sigma_\beta}{r}
\right)^{12}$;
$\alpha$ and $\beta$ are indexes that specify the particle species
($\alpha,\beta \in [A,B]$).
The potential is cut and shifted at $r_{\alpha \beta}=1.5 (\sigma_\alpha +
\sigma_\beta)$ as usually done
in Molecular Dynamics (MD) simulations \cite{allen}.
The small size of the system is mandatory in order to obtain the large
fluctuations of entropy production
needed to test Eq.~\ref{FT}.
A common choice for the particle radii is $\sigma_A/\sigma_B=1.2$
\cite{ParisiBMSS}.
All the quantities are then reported in units of
$m$, $\epsilon$, and the ``effective radius''
$\sigma_0^3 = [(2\sigma_A)^3 + 2 (\sigma_A+\sigma_B)^3 + (2\sigma_B)^3]/4$.
The particles are confined in a cubic box, at
density $\rho = 1$ , with periodic boundary condition adapted to the
presence of a shear flow. The SLLOD equations are integrated via a standard
velocity-Verlet algorithm that
approximates the exact equation of motion up to $o(dt^3)$ \cite{allen}.
The integration step is chosen to be $dt=0.002$ in order to have a very good
energy (or temperature) conservation over long times.

Three very long simulation runs ($10^9$~MD~steps, corresponding to $t=2 \cdot
10^6$) have been
performed in order to have good statistics also for $\tau \sim 10^5$~MD~steps:
the first one at equilibrium ($\gamma=0$) in the isokinetic ensemble at
$T=0.5$, the second one
in the same ensemble at the same temperature with $\gamma=0.05$, and the third
one in the
isoenergetic ensemble with $E=3.3$, that corresponds to $\langle T(p) \rangle
\sim 0.5$, and
$\gamma=0.05$.

\section{Conservative forces}

\begin{figure}[t]
\centering
\includegraphics[width=.70\textwidth,angle=0]{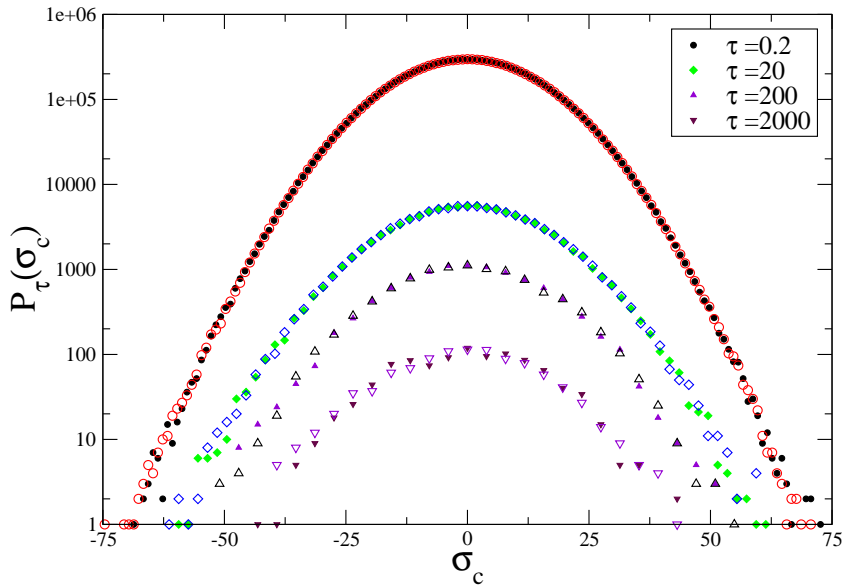}
\caption{Probability distribution functions of $\sigma_c$ for different values
of $\tau$ in equilibrium
and for $\gamma=0.05$: for each value of $\tau$, filled symbols correspond to
$\gamma=0$, open symbols
to $\gamma=0.05$. On the $y$ axis the number of counts in each bin of the
histogram is reported.
A Gaussian distribution (non reported) perfectly describes all the curves in
the range of fluctuations
accessible to our simulation. The width of the distribution does not depend on
$\tau$. }
\label{fig_1}
\end{figure}

First, we study the PDF of $\dot{\sigma}_c(t)$ in the isokinetic ensemble. In
Fig.~\ref{fig_1} the function
$P_\tau(\sigma_c)$
is reported for different values of $\tau$ (from $0.2$ to $2000$) in
equilibrium and for $\gamma=0.05$.
The two sets of distributions are observed to coincide over the whole time
range accessible to our simulations.
Thus $P_\tau(\sigma_c)$ is $\gamma$-independent and statement {\it ii)} is
verified.
At any time the distributions are well described by a Gaussian form
\beq
P_\tau(\sigma_c) \propto \exp \left[ -\frac{(\sigma_c - m_c(\tau))^2}{2
S^2_c(\tau)} \right]
\eeq
in the range of values of $\sigma_c$ accessible to our simulation.
From the fit of the data reported in Fig.~\ref{fig_1}, we find that - within
the statistical accuracy -
the mean value of $\sigma_c$ is vanishing, as expected (see the discussion
after Eq.~\ref{sigmaisoK}).
Moreover, recalling that
\beq
\sigma_c = \int_0^\tau \ dt \ \dot{\sigma}_c(t) = -\frac{1}{T}\int_0^\tau \ dt
\frac{dV(q)}{dt} =
\frac{V(0)-V(\tau)}{T} \ ,
\eeq
we find
\beq
\label{fluttcons}
S^2_c(\tau) = \langle \sigma_c^2 \rangle = 2 \frac{\langle V^2 \rangle -
\langle V(\tau) V(0) \rangle}{T^2} \sim 2 \frac{ \langle V^2 \rangle -  \langle
V \rangle^2}{T^2} \ ,
\eeq
where the last equality holds for large $\tau$. In our simulation $S_c^2(\tau)$
is observed to be
$\tau$-independent and equal to $2 ( \langle V^2 \rangle -  \langle V
\rangle^2)/T^2$
(i.e. to the fluctuations of the potential energy) in the whole investigated
$\tau$ range.

\section{Dissipative forces}
\label{sec:dissipa}

\begin{figure}[t]
\centering
\includegraphics[width=.70\textwidth,angle=0]{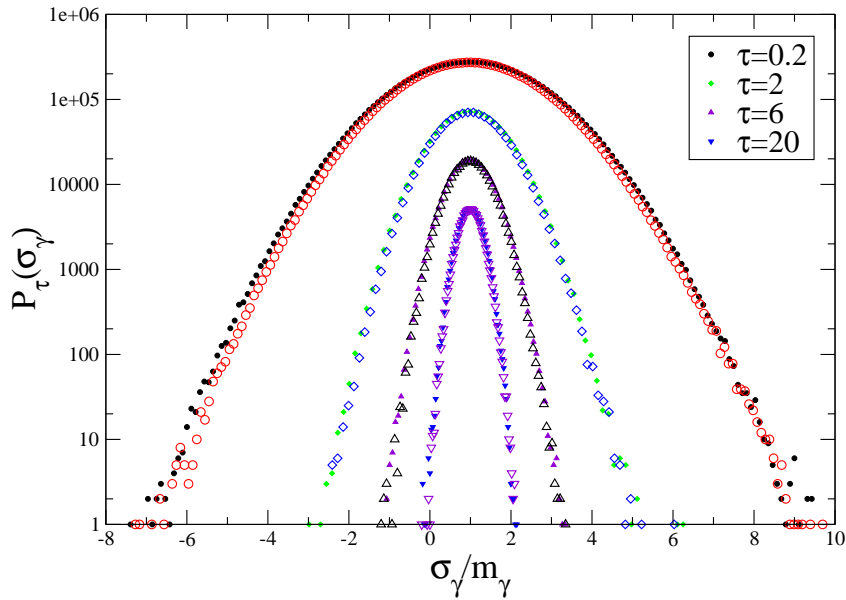}
\caption{$P_\tau(\sigma_\gamma)$ as a function of $\sigma_\gamma/m_\gamma$ at
$T=0.5$ and $\gamma=0.05$:
for each value of $\tau$ filled symbols correspond to the isokinetic ensemble,
open symbols correspond
to the isoenergetic ensemble. Ensemble equivalence holds for
this quantity for any value of $\tau$. The distributions are Gaussian over the
whole accessible range.}
\label{fig_2}
\end{figure}

\begin{figure}[ht]
\centering
\includegraphics[width=.70\textwidth,angle=0]{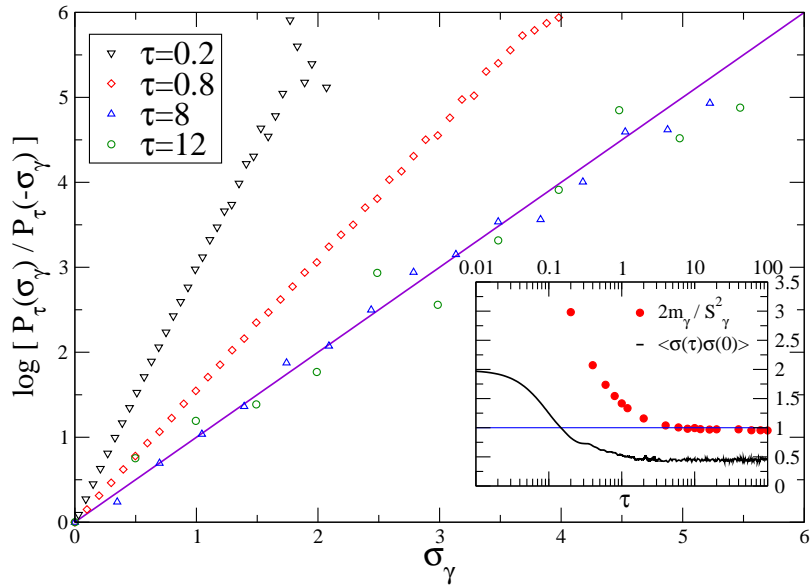}
\caption{$\log P_\tau(\sigma_\gamma)/P_\tau(-\sigma_\gamma)$ as a function of
$\sigma_\gamma$ for different
values of $\tau$ in the isokinetic ensemble. The FT predicts the slope to be
$1$ (continuous line) and is
verified for $\tau \geq 8$. In the inset $2 m_\gamma/S^2_\gamma$ (the slope of
the lines)
is reported as a function of $\tau$ together with the stress autocorrelation
function (rescaled by an
arbitrary factor).}
\label{fig_3}
\end{figure}

We turn now to the analysis of the PDF of $\dot{\sigma}_\gamma(t)$, the entropy
production due to the work of the
dissipative forces. This quantity is reported in Fig.~\ref{fig_2} for $T=0.5$
and $\gamma=0.05$ in both the
isokinetic and isoenergetic ensemble. The PDFs in the two ensembles are found
to coincide within the statistical
uncertainties: thus, {\it ensemble equivalence holds for the fluctuations of
$\dot{\sigma}_\gamma$ at any $\tau$},
proving the validity of statement {\it iii)}. \\
In the isoenergetic ensemble, the distribution $P_\tau(\sigma_\gamma)$ has been
shown to verify the FR
\cite{ECM}, so we expect that the FR is also verified in the isokinetic
ensemble for $\tau$ long enough.
In Fig.~\ref{fig_3} we report $\log P_\tau (\sigma_\gamma)/P_\tau
(-\sigma_\gamma)$ as a function of
$\sigma_\gamma$ calculated in the isokinetic ensemble.
The FR predicts that the curve must be a straight line with slope $1$. From
Fig.~\ref{fig_3} we see
that while for short $\tau$ values $\log P_\tau (\sigma_\gamma)/P_\tau
(-\sigma_\gamma)$ appears to be linear but
has a slope different from one, on increasing $\tau$ the slope tends toward one
and the FR is indeed verified
for $\tau \gtrsim 8$. \\
From Fig.~\ref{fig_2} we can also observe that, similarly to that of
$\dot{\sigma}_c(t)$,
the PDF of $\dot{\sigma}_\gamma(t)$
is Gaussian over a very wide range. The same behavior has been observed and
discussed in previous works
\cite{BGG,BCL,GP}.
For a Gaussian distribution,
$P_\tau(\sigma_\gamma) \propto \exp \left[ -\frac{(\sigma_\gamma -
m_\gamma(\tau))^2}{2 S^2_\gamma(\tau)} \right]$,
the FR can be expressed as a relation between the mean value and the variance:
\beq
\label{gaussian}
2 m_\gamma = S^2_\gamma \ .
\eeq
In the inset of Fig.~\ref{fig_3} the quantity $2 m_\gamma / S^2_\gamma$ is
reported as a function of $\tau$.
For $\tau \lesssim 20$ it coincides with the slope of $\log P_\tau
(\sigma_\gamma)/P_\tau (-\sigma_\gamma)$ as a
function of $\sigma_\gamma$ as derived from the main panel of Fig.~\ref{fig_3}.
For $\tau \gtrsim 20$ in our simulation it is not possible to observe negative
values of
$\sigma_\gamma$ and the function $\log P_\tau (\sigma_\gamma)/P_\tau
(-\sigma_\gamma)$
cannot be evaluated. However, the quantity $2 m_\gamma/S^2_\gamma$ can be
calculated also in absence
of negative values of $\sigma_\gamma$ and is found to be equal to one within
the statistical accuracy.
In the inset of Fig.~\ref{fig_3} the autocorrelation function of the entropy
production
$\langle \dot\sigma_\gamma(\tau) \dot\sigma_\gamma(0) \rangle$ is also reported
as a function of $\tau$.
This quantity, from Eq.~(\ref{sigmagamma}), is
proportional to the stress autocorrelation function $C(\tau)=\langle
P_{yx}(\tau) P_{yx}(0) \rangle$.
By defining the relaxation time of stress fluctuations $\tau_\alpha$ as
$C(\tau_\alpha)-C(\infty)=\frac{1}{2}[C(0)-C(\infty)]$, for the values of
$\gamma$ and $T$ analyzed here
we have $\tau_\alpha \sim 0.1$.
From the inset of Fig.~\ref{fig_3} we note that the FR is verified for $\tau
\gtrsim 10^2 \tau_\alpha$,
which is statement {\it iv)}.

\section{Total phase space contraction rate}

\begin{figure}[ht]
\centering
\includegraphics[width=.70\textwidth,angle=0]{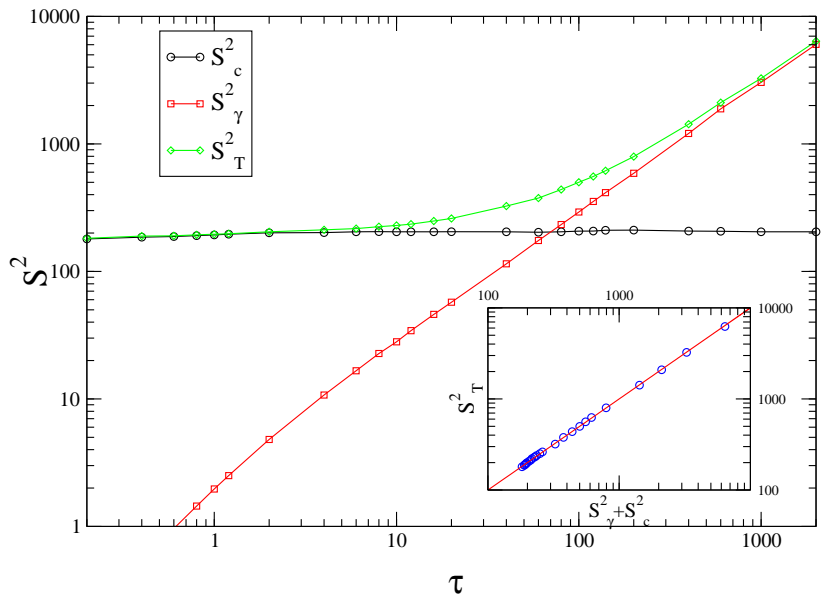}
\caption{Variances of $\sigma_c$, $\sigma_\gamma$ and $\sigma_T$ as a function
of $\tau$. In the inset,
$S^2_T$ is reported as a function of $S^2_c+S^2_\gamma$; the data agree
perfectly with the straight line
of slope 1: then, $\sigma_\gamma$ and $\sigma_c$ are statistically independent.
The fluctuations of $\sigma_T=\sigma_\gamma+\sigma_c$ are dominated
at short times by $\sigma_c$ and at long times by $\sigma_\gamma$.}
\label{fig_4}
\end{figure}

To conclude the analysis of the simulation data, we will now discuss the
behavior of the fluctuations
of the total phase space contraction rate, given by the sum of the two
previously discussed contributions.
Its PDF (not shown here) is also well described by a Gaussian distribution
$P_\tau(\sigma_T) \propto \exp \left[ -\frac{(\sigma_T - m_T(\tau))^2}{2
S^2_T(\tau)} \right]$
in the range of $\sigma_T$ accessible to our simulation.
To relate the distribution of $\sigma_T$ to the ones of $\sigma_c$ and
$\sigma_\gamma$, it is
sufficient to discuss the behavior of the corresponding variances.
In fact, remembering that $m_c=0$, $m_\gamma=m_T = 1.39 \tau$ and all the
distributions are Gaussian in
the observed range, the three distributions are fully specified by their
variances.
In Fig.~\ref{fig_4} we report the variances $S^2_c$, $S^2_\gamma$ and
$S^2_T$ as a function of $\tau$. In the inset, $S^2_T$ is reported as a
function of $S^2_c+S^2_\gamma$
parametrically in $\tau$.
From Fig.~\ref{fig_4} we deduce that $\sigma_c$ and $\sigma_\gamma$ are
statistically independent, as the
variance of their sum is the sum of their variances, thus verifying statement
{\it i)}.
Then, as $\sigma_T=\sigma_c+\sigma_\gamma$, its distribution is the convolution
of the distributions of
$\sigma_c$ and $\sigma_\gamma$.
We observe also that $S^2_c$ is time-independent, while $S^2_\gamma = 2
m_\gamma = 2.78 \tau$ as predicted
by the FR. Then, the fluctuations of the total entropy production are dominated
by the conservative part at
short times and by the dissipative part at long times. We have
\beq
\begin{cases}
&P_\tau(\sigma_T) \sim P_\tau(\sigma_c) \hspace{1cm} \text{for small $\tau$
($\lesssim 5$)} \\
&P_\tau(\sigma_T) \sim P_\tau(\sigma_\gamma) \hspace{1cm} \text{for large
$\tau$ ($\gtrsim 1000$)}
\end{cases}
\eeq
This implies that the FR, being verified for $P_\tau(\sigma_\gamma)$, is
verified also for the PDF of the
total entropy production $\sigma_T$, but for much longer times ($\tau \gtrsim
1000 \sim 10^4 \tau_\alpha$),
and verifies statement {\it v)}.
Note that in our simulation it is not possible to directly check the validity
of the FR in the asymptotic
regime for $\sigma_T$ because it is not possible to observe negative values of
$\sigma_T$ over so long times,
so that the validity of the FR has to
be deduced assuming a Gaussian form for the distribution $P_\tau(\sigma_T)$ and
checking that the ratio
$2 m_T/S^2_T$ is equal to one.
However, by looking to the fluctuations of the dissipative part alone, the
asymptotic regime is easily reached
and a direct verification of the FR in the isokinetic ensemble is possible.
Note also that the asymptotic regime for $\sigma_T$ corresponds to a regime in
which the fluctuations due to the dissipative part are dominant and $\sigma_T
\sim \sigma_\gamma$.

\section{Fluctuation Relation for a Gaussian distribution}

It is interesting to discuss briefly the implication of the FR if the
distribution
$P_\tau(\sigma)$ is Gaussian \cite{GAL,BGG}. In this case, the FR is equivalent
to Eq.~\ref{gaussian},
and we can easily rewrite, using Eq.~\ref{sigmagamma} and time-translation
invariance:
\beq
\begin{split}
&m_\gamma(\tau)=-\frac{\gamma}{T} \langle P_{yx} \rangle \tau \\
&S^2_\gamma(\tau)=\frac{\gamma^2}{T^2} \int_0^\tau dt \int_0^\tau dt' \ \langle
P_{yx}(t) P_{yx}(t') \rangle =
\frac{2 \gamma^2 \tau}{T^2} \int_0^\tau dt \ C(t)
\end{split}
\eeq
Then, Eq.~\ref{gaussian} can be written in the following form
\beq
\langle P_{yx} \rangle = -  \frac{\gamma}{T} \int_0^\tau dt \ C(t)
\eeq
or, defining the viscosity $\eta = -\frac{\langle P_{yx} \rangle}{\gamma}$,
\beq
\eta = \frac{1}{T} \int_0^\tau dt \ C(t)
\eeq
This (for $\tau \rightarrow \infty$) is an example of the well known Green-Kubo
relation for the transport coefficients,
that is strictly valid only in the $\gamma \rightarrow 0$ limit.
Thus, {\it if the distribution $P_\tau(\sigma_\gamma)$ is a Gaussian,
the FR implies the validity of the Green-Kubo relation for the viscosity also
at finite $\gamma$.}

\section{Conclusions}

We discussed a simple model of a driven non-conservative system, i.e. a system
in which the phase space
volume is not conserved in equilibrium: a sheared liquid in the isokinetic
ensemble.

We have shown that the total phase space contraction rate is the sum of two
statistically independent
contributions, defined in Eq.~\ref{sigmaisoK}: the first,
$\dot{\sigma}_\gamma$, is due to the work
of the dissipative forces, and is the microscopic equivalent of the
thermodynamic definition of entropy
production given in Eq.~\ref{DEFsigma}.
Its PDF is found to be the same in the two considered ensembles for any value
of
$\tau$, and verifies the FR in the large $\tau$ limit, $\tau \gtrsim 10^2
\tau_\alpha$.
The second contribution is $\gamma$-independent, has zero average and is
negligible in the
(very) large $\tau$ limit, $\tau \gtrsim 10^4 \tau_\alpha$.

Then, we can conclude that: \\
{\it a)} The total entropy production can be identified, in the isokinetic
ensemble, with the total phase space
contraction rate $\dot{\sigma}_T$. The equivalence of isokinetic and
isoenergetic ensembles holds, as
conjectured in \cite{gal1,gal2,gal_stat,gal_fluid}.
However, very large values of $\tau$, such that the contribution of
$\dot{\sigma}_c$ can be neglected,
have to be reached. \\
{\it b)} If one looks at {\it that part of the phase space contraction rate
that vanishes in the equilibrium limit},
namely $\dot\sigma_\gamma$, equivalence holds for any $\tau$ and the FR is
found to be verified at shorter times
(by a factor $10^2$). \\
Obviously the first definition of entropy production rate has general validity
\cite{gal2}, while the second one
has to be discussed case by case by identifying the ``relevant'' part of the
total phase space contraction rate.
However, the second definition turns out to be very useful on a practical
ground, because the very large $\tau$ values
needed to observe the validity of the FR for the total phase space contraction
rate are difficult to be reached
in computer simulations. Also, this observation is relevant for experiments on
real systems: indeed,
equilibrium fluctuations of entropy production (analogous to $\dot{\sigma}_c$)
are always present in
real system in contact with a thermal bath, and it is impossible to separate
the two contributions as in a
numerical experiment. In planning experiments, one has then to check carefully
that the
time scales involved are such that the contribution analogous to $\dot\sigma_c$
is negligible.

\hspace{1cm}

It is a pleasure to thank Giovanni Gallavotti for a careful reading of the
manuscript and for
interesting suggestions and comments, and Alessandro Giuliani for useful
discussions
and encouragement.

%%%%%%%%%%%%%%%%%%%%%%%%%%%%%%%%%%%%%%%%%%%%%%%%%%%%%%%%%%%%%%%%%%%%%%%%%%%
%                             REFERENCES
%%%%%%%%%%%%%%%%%%%%%%%%%%%%%%%%%%%%%%%%%%%%%%%%%%%%%%%%%%%%%%%%%%%%%%%%%%%

\end{document}